\documentclass[twocolumn,showpacs,preprintnumbers,amsmath,amssymb]{revtex4}
\usepackage{graphicx}
\usepackage{rotating}
\usepackage{dcolumn}
\usepackage{bm}
\begin{document}

%\preprint{NLP/0401-SJTU}

\title{Vortices, circumfluence, symmetry groups \\ and Darboux transformations of the Euler equations}
\author{S. Y. Lou$^{1,2}$, X. Y. Tang$^{1,2}$, M. Jia$^{2}$ and F. Huang$^{1,2,3}$}
\affiliation{$^{1}$Department of Physics, Shanghai Jiao Tong
University,Shanghai, 200030, China\\
$^{2}$Center of Nonlinear Science, Ningbo University, Ningbo,
315211, China\\
$^{3}$Department of Marine Meteorology, Ocean University of China,
Qingdao 266003, China}
\date{\today}% It is always \today, today,

\begin{abstract}
The Euler equation (EE) is one of the basic equations in many
physical fields such as the fluids, plasmas, condense matters,
astrophysics, oceanic and atmospheric dynamics. A new symmetry
group theorem of the two dimensional EE is obtained via a simple
direct method and the theorem is used to find \em exact analytical
\rm vortex and circumfluence solutions. Some types of Darboux
transformations (DTs) for the both two and three dimensional EEs
are obtained for \em arbitrary spectral parameters \rm which
indicates that the EEs are integrable and the Navier-Stockes (NS)
equations with large Renoyed number are nearly integrable, i.e,
they are singular perturbations of the integrable EEs. The
possibility of the vortex and circumfluence solutions to
approximately explain the tropical cyclones (TCs), especially, the
Hurricane Andrew 1992, is discussed.
\end{abstract}

% PACS, the Physics and Astronomy Classification Scheme.

\maketitle

{\em 1. Introduction.} In fluid physics, there are various
important open problems. One of the most important of them is the
existence and smoothness problem of the NS equation which is the
basic equation and the starting point of all the problems in fluid
physics \cite{NS}. Due to its importance and difficulty, the
problem is listed as one of The Millennium Problems of the 21st
century \cite{clay}.

One of the most significant recent development related to this
problem may be the discovery of the weak Lax pairs of the two and
three dimensional EEs which are the limit cases of the NS equation
for the large Renoyed number \cite{Li}. That means the EEs are
integrable and thus the NS equation with large Renoyed number is
nearly integrable, the singular perturbation of integrable models!

The (3+1)-dimensional EE
\begin{eqnarray}
E_1\equiv \vec{\omega}_t+(\vec{u}\cdot\nabla)
\vec{\omega}-(\vec{\omega}\cdot\nabla) \vec{u} =0,\
\vec{\omega}=\nabla \times \vec{u},\ \label{E}
\end{eqnarray}
with $ \nabla \cdot \vec{u}=0$ is the original springboard for
investigating the incompressible inviscid fluid. In the equation
system \eqref{E}, $\vec{\omega}\equiv\{\omega_1,\ \omega_2,\
\omega_3\}$ is the vorticity and $\vec{u}\equiv\{u_1,\ u_2,\
u_3\}$ is the velocity of the fluid. In (2+1)-dimensional case,
the EE is simplified to
($[\psi,\omega]\equiv\psi_x\omega_y-\psi_y\omega_x$)
\begin{eqnarray}
E\equiv\omega_t+[\psi,\omega]=0,\ \omega=\psi_{xx}+\psi_{yy},\
\label{e}
\end{eqnarray}
where the velocity $\vec{u}=\{u_1,\ u_2\}$ is linked to the stream
function $\psi$ by $u_1=-\psi_{y},\ u_2=\psi_{x}.$

The EEs are important not only in fluid physics \cite{fluid} but
also in many other physical fields such as the plasma physics
\cite{plasma}, oceanography \cite{ocean}, atmospheric dynamics
\cite{gas}, superfluid and superconductivity \cite{super},
cosmography and astrophysics \cite{astro}, statistical physics
\cite{sta}, field and particle physics\cite{particle} and condense
matters such as the Bose-Einstein condensation \cite{bose}, the
crystal liquid \cite{crystal} and the liquid metallic hydrogen
\cite{H} etc.

There are innumerable papers on the EEs in literature due to EE
\eqref{E} or \eqref{e} is just the beginning point to study
various physical problems. However, most of them only treat the
equation approximately or numerically because the weakly Lax
integrability of the model has just been revealed most recently by
Charles Li \cite{Li}. So there is few results on the \em exact \rm
and \em analytic \rm solutions of the EEs.

It is known that to find some exact and analytic results of a
physical system, Lie group theory is one of the most effective
methods. Nevertheless, even for a mathematician, to find the
symmetry group of a given nonlinear system is still very difficult
especially for non-Lie and non-local symmetry groups. So it is
significant for a physicist to find a \em simple \rm method to get
\em more general \rm symmetry groups of nonlinear systems without
use of the complicated group theory.

Vortices and circumfluence are most general observations in some
physical fields especially for fluid systems. In our knowledge,
there is little exact analytic understanding on vortices and
circumfluence though there are rich vortex structures for
different physical systems. If one can find the full symmetry
group of the EE, one may get some kinds of exact vortex and
circumfluence solutions from some simple trivial or periodic seed
solutions.

Though the (weak) Lax pairs of the EEs were found in both two and
three dimensions four years ago, little further understanding of
the exact solutions has been obtained from their Lax pairs. For
the two-dimensional EE, one special type of DT with zero spectral
parameter had also been given by Li \cite{Li}. In order to get
nontrivial and as many as possible results from DT, one has to
find DT with \em nonzero \rm spectral parameter which is still
failed for both two and three dimensional situations.

In this Letter, we firstly establish a simple direct method to
find a general group transformation theorem for the two
dimensional EE. Then the theorem is used to find a quite general
symmetric vortex and/or circumfluence solution with some arbitrary
functions starting from a quite trivial periodic solution (Rosbby
wave). Some velocity field plots related to vortices and
circumfluence are given. The exact vortices and the circumfluence
solutions are used to explain TC eye, track and the relation
between track and the background wind. Some DTs with \em arbitrary
spectral parameters \rm for two and three dimensional EEs are
explicitly given.

{\em 2. Space-time transformation group of the two dimensional
EE.} In the traditional theory, to find the Lie symmetry group of
a given nonlinear physical system, one has to firstly find its Lie
symmetry algebra and then use the First Fundamental Theorem to
solve an ``initial" problem. If one utilize the standard Lie group
theory to study the symmetry group of the two-dimensional EE, it
is easy to find that the only possible symmetry transformations
are: the arbitrary time dependent space and stream translations,
constant time translation, space rotation and
%, time dependent scaling transformation and space dependent
scaling transformations. The details on the traditional Lie point
symmetries can be found in the literature, say, \cite{HF}.

Recently, it is found that for simplicity and to find \em more
general \rm symmetry groups, one may use some types of new simple
direct method especially for Lax integrable models without any
group theory \cite{group_JPA,CPL}.

For the two-dimensional EE, its weak Lax pair reads
\begin{eqnarray}
&& \omega_x\phi_{y}-\omega_y\phi_{x}=\lambda\phi,\
\phi_t+\psi_x\phi_y-\psi_y\phi_x=0,  \label{Lax1}
\end{eqnarray}
where $\lambda$ is the spectral parameter. That means the
compatibility condition of \eqref{Lax1} is just the EE \eqref{e}.
We say the Lax pair is weak because that $\eqref{E}$ is a
compatibility condition only under the meaning that $[E,\ \phi]=0$
\cite{Li}.

It is not difficult to understand that the symmetry group of the
EE can be obtained by using the following gauge transformation
companied with the space time transformation for the spectral
function $\phi$
\begin{eqnarray}
\phi \rightarrow g \phi'(\xi,\ \eta,\ \tau)\equiv g\phi',
\label{trans}
\end{eqnarray}
where $g,\ \xi,\ \eta$ and $\tau$ are undetermined functions of
$\{x,\ y,\ t\}$. The function $\phi'$ in \eqref{trans} should
satisfy the same Lax pair \eqref{Lax1} but with different
independent variables, i.e.,
\begin{eqnarray}
\omega'_{\xi}\phi'_{\eta}-\omega'_{\eta}\phi'_{\xi}=\lambda'\phi',\
\phi'_{\tau}+\psi'_{\xi}\phi'_{\eta}-\psi'_{\eta}\phi'_{\xi}=0,
\label{Lax2}
\end{eqnarray}
where $\{\omega',\ \psi'\}$ is also a solution of the EE \eqref{e}
under the transformation $\{x,\ y,\ t,\ \lambda\}\rightarrow
\{\xi,\ \eta,\ \tau,\ \lambda'\}$. Substituting the gauge
transformation \eqref{trans} into the Lax pair \eqref{Lax1},
eliminating $\phi'_{\eta}$ and $\phi'_{\tau}$ by means of
\eqref{Lax2} and vanishing the coefficients of $\phi'_\xi$ and
$\phi'$ of the resulting equations, we have the determining
equations for the functions $\xi,\ \eta,\ \tau,\ \psi',\ \omega'$
and $g$: {\small
\begin{eqnarray}
&&\{([\tau, \omega]\psi'_\xi+[\omega,
\eta])\lambda'-\lambda\omega'_\xi\}g+[\omega,
g]\omega'_\xi=0,\nonumber\\
&&\{\eta_t+[\psi,  \eta]-(\tau_t+[\psi, \tau])
\psi'_\xi\}\lambda'g+(g_t+[\psi, g])\omega'_\xi=0,\nonumber\\
&&(\tau_t+[\psi, \tau])\omega'_\tau+(\eta_t+[\psi,
\eta])\omega'_\eta+(\xi_t+[\psi, \xi])\omega'_\xi=0,\nonumber\\
&&[\tau,  \omega]\omega'_\tau+[\omega, \eta]\omega'_\eta+[\omega,
\xi]\omega'_\xi=0.\label{det}
\end{eqnarray}}
 It is clear that there are abundant interesting exact
solutions because the determining equation system \eqref{det} is
under-determined. Here we just write down one special solution
theorem of the two dimensional EE.

{\bf Theorem 1.} {\em If $\{\omega(x,\ y,\ t),\ \psi(x,\ y,\ t)\}$
is a known solution of the two dimensional EE, so is
$\{\omega'\equiv F(\omega(\xi,\ \eta,\ \tau),\ t)\equiv F,\
\psi'\equiv G(\omega(\xi,\ \eta,\ \tau),\ t)+f\equiv G+f\}$ with
$f=f(x,\ y,\ t),\ \xi=\xi(x,\ y,\ t),\ \eta=\eta(x,\ y,\ t)$ and
$\tau\equiv \tau(t)$ are functions of the indicated variables
while the six functions $\{f,\ \xi,\ \eta,\ \tau,\ F,\ G\}$ need
to satisfy two constrained equations
\begin{eqnarray}
&& F=G_{\omega\omega}(\omega_x^2+\omega_y^2)+G_\omega(\omega_{xx}+\omega_{yy})+f_{xx}+f_{yy},\nonumber\\
&& f_x\omega_y-\omega_xf_y+\omega_t=0,\label{th}
\end{eqnarray}
where $G_\omega =\partial G/\partial \omega$ and $\omega_x,\
\omega_y$ and $\omega_t$ are total derivatives, say,
$\omega_t=\omega_\xi\xi_t+\omega_\eta\eta_t+\omega_\tau\tau_t.$}

Starting from some simple seed solutions, one can obtain many
 physically interesting solutions from Theorem 1.
For instance, if we take the seed solution as the usual Rossby
wave solution
\begin{eqnarray}
&& \psi=ax-by+c\cos(kx+ly+(ka+bl)t),\nonumber\\
&& \omega=-(k^2+l^2)c\cos(kx+ly+(ka+bl)t),\label{seed}
\end{eqnarray}
then we can get the following new general vortex and/or
circumfluence solution ($r\equiv (x-x_0)^2+(y-y_0)^2$):
\begin{eqnarray}
\psi'&=&y_{0t}x-x_{0t}y+F_1\ln r+F_2
\nonumber\\
&&-\frac{1}2h_{0t}\tan^{-1}\frac{x-x_0}{y-y_0}+\frac14\int \frac{F(r+h_0)}r{\rm d}r,\nonumber\\
\omega'&=&F_r(r+h_0),\label{vortex}
\end{eqnarray}
where $x_0,\ y_0,\ h_0,\ F_1$ and $F_2$ are arbitrary functions of
$t$ and $F\equiv F(r+h_0)$ is an arbitrary function of $r+h_0$.

Because of the intrusion of the many arbitrary functions into the
exact solution \eqref{vortex}, one is able to find various vortex
and circumfluence structures by selecting the arbitrary functions
in different ways. Here are some special
examples for $h_0=F_1=F_2=0$.\\
{\em 3. Lump-type vortices.} If we take the function $F(r)$ as a
rational solution of $r$ , $
F(r)=\frac{\sum_{i=0}^Na_ir^i}{\sum_{i=0}^Nb_ir^i}\equiv
\frac{P(r)}{Q(r)},$ with the conditions $b_N\neq0$ and $Q(r)\neq
0$ for all $r\geq 0$, then the solution \eqref{vortex} becomes an
analytical lump-type vortex and/or circumfluence solution for the
velocity field. For instance, Fig. 1a shows a typical lump-type
vortex solution for the velocity field with the selection, $
F(r)=-{4r^2}{(1+r)^{-2}}$, at times for $x_0=y_0=0$. The related
velocity possesses the rational form{\small
\begin{equation}
u={2(y-y_0)r}{(1+r)^{-2}},\ v={2(x_0-x)r}{(1+r)^{-2}}.
\label{lump1uv}
\end{equation}}
\input epsf
     \begin{figure}
     \epsfxsize=7.5cm\epsfysize=3.5cm\epsfbox{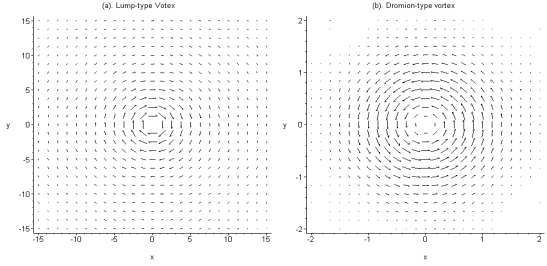}
\caption{(a). A field plot of the typical lump-type vortex given
by \eqref{lump1uv} with $x_0=y_0=0$. (b). A field plot of the
dromion-type vortex with the selection \eqref{dromionuv} for
$x_0=y_0=0$.}
     \end{figure}
{\em 4. Dromion-type vortices}. If we take the function $F(r)$ as
a rational function of $r$ multiplied by an exponentially decayed
factor, say, $\exp(-r)$, then \eqref{vortex} becomes an analytical
dromion-type vortex and/or a circumfluence solution. Fig. 1b shows
a typical dromion-type vortex solution for the velocity field with
\begin{eqnarray}
F(r)=100re^{-r}, \label{dromionuv}
\end{eqnarray}
at times $t_i$ for $x_0(t_i)=y_0(t_i)=0$.

{\em 5. Ring soliton solutions and circumfluence}. In the recent
studies of (2+1)-dimensional nonlinear physics systems, some kinds
of ring soliton solutions have been found \cite{ring,ring1}. It is
interesting that the basin-type and/or plateau-type of soliton
solutions may be responsible for the circumfluence solution for
the fluid system described by EE. For instance, if we select the
function $F(r)$ possesses the property $\left. {d^i F(r)}/{d
r^i}\right|_{r=0}=0,\ i=0,\ 1,\ ...,\ n$ for $n\geq 2$, then
\eqref{vortex} expresses the circumfluence for the velocity field
and the basin-type or plateau type ring soliton for the stream
function. Fig. 2 is a special plot of \eqref{vortex} with the
selections $x_0=y_0=0$ and
\begin{eqnarray}
F(r)=-4{r^{2}e^{-r}}. \label{circulation}
\end{eqnarray}

Fig. 2a exhibits the circumfluence structure for the velocity
field related to the selection \eqref{circulation} and Fig. 2b
shows the basin type ring soliton shape for the stream function
$\psi$ with the same selections as Fig. 2a.
% while the related configuration of the voticity $\omega$ is shown in Fig. 3c.\\
\input epsf
\begin{figure}
\epsfxsize=7.5cm\epsfysize=3.5cm\epsfbox{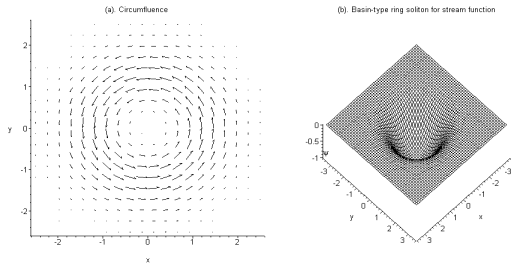}
\caption{(a). A field plot of the circumfluence with the selection
\eqref{circulation} and $x_0=y_0=0$ for the velocity $\{u,\ v\}$.
(b). A special basin type ring soliton for the stream function
$\psi$ related to (a).
%(c). Ring soliton structure for the vorticity $\omega$.
}
\end{figure}

{\em 6. Tropical TC track and background wind.} From the
expression of the exact solution \eqref{vortex} and the Figs.
1--2, we know that the vortices in physics may have quite rich
structures. Due to the richness of the solution structures and
wide applications of the vortex in various fields such as the
fluids, plasma, oceanic and atmospheric dynamics, cosmography,
astrophysics, condense matters etc \cite{fluid}--\cite{H}, the
results may be applied in all of these fields. For instance, in
the oceanic and atmospheric dynamics, the analytical solutions
expressed by \eqref{vortex} may be used to approximately describe
the TCs which possess increasing destructiveness over the past 30
years \cite{typhoon}. The relatively tranquil part, the center of
the circumfluence shown by Fig. 2 is responsible for the TC eye
\cite{eye}. Furthermore, the expression \eqref{vortex} also offers
a relation between the TC track given by $\{x_0(t),\ y_0(t)\}$ and
the strength of the background wind described by $\{x_{0t},\
y_{0t}\}$. Fig. 3 shows an example on the TC track (Fig. 3b) and
the related background wind field (Figs. 3c--3f). To compare with
real observations, the track of the Hurricane Andrew 1992 from 23
to 27 August is also plotted at Fig. 3a. From Fig.3 one can find
that the background wind lead to the change of the direction of
the TC track.
\input epsf
\begin{figure}
\epsfxsize=7.5cm\epsfysize=3.5cm\epsfbox{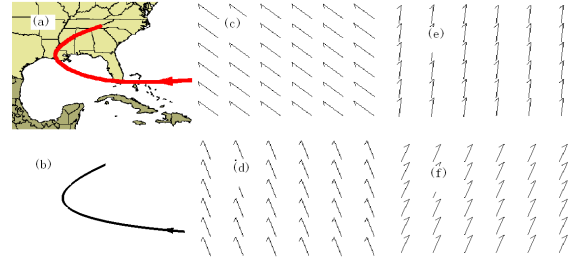} \caption{\small
(a) The real TC track of the Hurricane Andrew 1992 from 23 to 27
August. (b) The theoretical track given by $x_0=0.1(t-3.7)^2-3,\
y_0=0.1t^2-3$ from day 1 to day 5. (c)--(f) The background wind
fields related to the TC track (b) at day 2, 3, 4 and 5
respectively.}
\end{figure}

{\em 7. DTs of EEs.} In \cite{Li}, Li found a DT
 of EE \eqref{e} with Lax
pair \eqref{Lax1} for \em zero \rm spectral parameter. For the
three-dimensional EE, though two types of weak Lax pairs had also
been given by Li et al, there is no DT found even for zero
spectral parameter. Here we write down one new DT for the
two-dimensional EE \eqref{e} and two DT theorems for the
three-dimensional EE \eqref{E} with \em arbitrary spectral
parameters  \rm without any proof for lacking of
space.\\
 {\bf Theorem 2.} {\em If $\{\omega,\ \psi,\ \phi\}$ is
a solution of \eqref{e} and \eqref{Lax1} with the spectral
parameter $\lambda$, $G(f)$ being an arbitrary function of $f$
that is a given spectral function of \eqref{Lax1} under the
spectral parameter $\lambda_0$, then $\{\omega+q,\ \psi+p,\
G\phi\}$  with the spectral parameter $\lambda_1$ is also a
solution where $p$ and $q$ are determined by $q\equiv
p_{xx}+p_{yy}$ and
\begin{eqnarray}
[p,\ G\phi]=0,\ [q,\
G\phi]=(\lambda_1-\lambda)G\phi-\lambda_0G_f.\label{pq}
\end{eqnarray}}
It is interesting that if we take all the parameters $\lambda,\
\lambda_0$ and $\lambda_1$ as zero, the B\"acklund transformation
$\omega'=\omega+q,\ \psi'= \psi+p$ with \eqref{pq} is equivalent
to that obtained by Li \cite{Li}. To see this point more clearly,
one can write \eqref{pq} in an alternative form by eliminating
$\phi_y$ via Lax pair \eqref{Lax1}{\small
\begin{eqnarray}
&&\lambda_1\omega_x-[\lambda+\lambda_0f(\ln
G)_f](\omega+q)_x+[\omega,\ q][\ln (G\phi)]_x=0,\nonumber\\
&&[\omega+q,\ p][\ln (G\phi)]_x+\lambda_1p_x=0.\label{pq1}
\end{eqnarray}}
{\bf Theorem 3.} {\em If $\{\vec{\omega},\ \vec{u},\ {\phi}\}$ is
a solution of EE \eqref{E} and its scalar Lax pair \cite{Li}
$\{\vec{\omega}\cdot \nabla \phi =\lambda \phi,\
\phi_t+\vec{u}\cdot \nabla \phi=0\}$, so is
$\{\vec{\omega}+\vec{q},\ \vec{u} +\vec{v},\ G\phi\}$ with the
spectral parameter $\lambda_1$, where $G$ is an arbitrary spectral
function of $f$ under the spectral parameter $\lambda_0$ while
$\vec{v}$ and $\vec{q}$ are determined by $\phi G_f \vec{q}\cdot
\nabla f+G\vec{q}\cdot \nabla \phi+\lambda_0f\phi
G_f+(\lambda-\lambda_1)G\phi=0$ and $\phi G_f \vec{v}\cdot \nabla
f +G\vec{v}\cdot \nabla \phi=0.$}

 {\bf Theorem 4.} {\em If
$\{\vec{\omega},\ \vec{u},\ \vec{\phi}\}$ is a solution of EE
\eqref{E} and its vector Lax pair \cite{Li}$\{ \vec{\omega}\cdot
\nabla \vec{\phi}-\vec{\phi}\cdot \nabla \vec{\omega} =\lambda
\vec{\phi},\ \vec{\phi}_t+\vec{u}\cdot \nabla
\vec{\phi}-\vec{\phi}\cdot \nabla \vec{u}=0,\}$, so is
$\{\vec{\omega}+\vec{q},\ \vec{u} + \vec{v},\ \vec{f}\diamond
\vec{\phi}\}$ with the spectral parameter $\lambda_1$ and
$\vec{f}$ being a given spectral vector function under the
spectral parameter $\lambda_0$ while $\vec{v}$ and $\vec{q}$
determined by $(\vec{q}\cdot \nabla) (\vec{f}\diamond
\vec{\phi})-[(\vec{f}\diamond \vec{\phi})\cdot
\nabla]\vec{q}=(\lambda_1-\lambda-\lambda_0)\vec{f}\diamond
\vec{\phi}$ and $(\vec{v}\cdot \nabla) (\vec{f}\diamond
\vec{\phi})-[(\vec{f}\diamond \vec{\phi})\cdot \nabla]\vec{v}=0$,
where ``$\diamond$" can be any kinds of possible vector product
(any possible map from $R^3\times R^3 \rightarrow R^3$) as long as
the Lebnitz differential rule is valid, i.e., $(\vec{a}\diamond
\vec{b})_x=\vec{a}_x\diamond \vec{b}+\vec{a}\diamond \vec{b}_x$.}

The simplest example of ``$\diamond$" may be ``$\times$", the
usual cross product of vectors.

In principle, the repeated uses of the DT theorems 2--4 and some
known seed solutions, one may get infinitely many new solutions of
the two  and three dimensional EEs. In real applications, it is
still very difficult. For instance, though a special DT with zero
spectral parameter was given four years ago, no concrete new
solutions have been obtained from it up to now. Here we just offer
a special concrete example for the two-dimensional EE.

It is obvious that the constant flow $u_1=b,\ u_2=a$, i.e.,
$\psi=ax-by,\ \omega=0$, is a solution of the two-dimensional EE.
This trivial solution corresponds to $\lambda=\lambda_0=0$.
Starting from this simple seed, Theorem 2 leads to a quite general
but complicated solution with some arbitrary functions. It is
interesting that the usual Rossby wave \eqref{seed} is just a
special case if we take $\lambda_1=0$. That means the Rossby wave
\eqref{seed} can be obtained as a special case of the first step
DT with zero spectral parameters.

{\em 8. Summary and discussion.} The analytical and exact forms of
the vortices and circumfluence of the two dimensional fluid are
studied by means of the symmetry group theorem 1 and the DT
theorem 2 of the two dimensional EE. Starting from the trivial
constant flow, the periodic Rossby wave are obtained as a special
case of the first step DT. Using the group transformation theorem
to the Rossby wave, a solution with some arbitrary functions can
be obtained. The solution including many kinds of possible
vortices and circumfluence such as the lump type vortices, dromion
type vortices etc. The vortex and circumfluence solutions may be
applied in various physics fields mentioned in the introduction
section and the references \cite{fluid}--\cite{H}. Especially,
they can be applied to approximately explain some important
problems of TCs such as its eye, track and the relations between
track and the background wind.

The DTs with arbitrary spectral parameters are successfully
obtained. The existence of the DTs with \em arbitrary spectral
parameter \rm is important not only because infinitely many exact
solutions of the EEs can be obtained but also the integrability of
the EEs is confirmed. Hereafter the NS equations with large
Renoyed number which are related to the usual important cases can
be studied by means of the singular perturbations of the
integrable ones. Consequently, the more general physical problems
can be studied better. For instance, for the TC problems we can
use the exact solutions of the EEs to study their steady areas
near the eyes and the singular perturbation to investigate their
chaotic parts, the areas around the eyes.

Because of the importance of the EEs and the NS system and their
wide applications, the more about the models, their DTs and the
exact solutions given in this Letter are worthy of further study.

Lou is in debt to thank Profs. Y. S. Li, X. B. Hu and Q. P. Liu
for their helpful discussions. The work was supported by the
National Natural Science Foundation of China (No. 90203001, No.
10475055 and No. 40305009).


\begin{thebibliography}{999}
\bibitem{NS}D. Sundkvist, V. Krasnoselskikh, P. K. Shukla, A. Vaivads, M. Andr\'e, S. Buchert and H. R\`eme, Nature,
\bf 436 \rm 825 (2005); G. Pedrizzetti, Phys. Rev. Lett. \bf 94
\rm 194502 (2005).
%; A. Groisman, S. R. Quake, Phys. Rev. Lett. \bf 92 \rm 094501 (2004).
%; V. L. Saveliev, M. A. Gorokhovski, Phys. Rev. E \bf 72 \rm 016302 (2005).
\bibitem{clay} C. L. Fefferman,
%Existence and smoothness of Navier-Stokes equation,
http://www.claymath.org /millennium /Navier-Stokes \_Equations
/Official\_Problem\_ Description. pdf (2000).
\bibitem{Li}Y. G. Li, J. Math. Phys. \bf 42\rm, 3552 (2001);
%Acta Appl. Math. \bf 77\rm 181 (2003);
Y. G. Li and A. V. Yurov, Stud. Appl. Math. \bf 111\rm 101
(2003).
%; Y. C. Li and R. Shvidkoy, J. Math. Anal. Appl. \bf 292\rm 311 (2004).
\bibitem{fluid}
%A. J. Majda and A. L. Bertozzi, Vorticity and Incompressible Flow (Cambridge University Press, Cambridge, England, 2001);
P. H. Chavanis and J. Sommeria, Phys. Rev. Lett.
\bf 78\rm, 3302 (1997); P. H. Chavanis, \em ibid, \bf 84\rm,
 5512 (2000).
 %; P. B. Weichman and M. Petrich, \em ibid, \bf 86\rm,  1761 (2001).
\bibitem{plasma}
%B. N. Kuvshinov, F. Pegoraro and T. J. Schep, Phys. Lett. A \bf 191, \rm 296 (1994);
E. Cafaro et al Phys. Rev. Lett. \bf 80\rm, 4430 (1998);
%D. Grasso, F. Califano, F. Pegoraro and F. Porcelli, \em ibid \bf
%86\rm, 5051 (2001);
D. Del Sarto, F. Califano and F. Pegoraro, \em
ibid \bf 91\rm, 235001 (2003).
%; B. N. Kuvshinov, F. Pegoraro, J. Rem and T. J. Schep, Phys. Plasmas \bf 6\rm, 713 (1999); J.
%Bergmans and T. J. Schep, Phys. Rev. Lett. \bf 87\rm, 195002
%(2001); J. L. Thiffeault and P. J. Morrison, Physica D \bf 136\rm, 205 (2002).
\bibitem{ocean}J. Marshall, A. Adcroft, C. Hill, L. Perelman and
C. Heisey, J. Geophys. Res. Oceans (C3) \bf 102 \rm, 5753 (1997);
V. M. Canuto and M. S. Dubovikov, Ocean Modelling, \bf 8 \rm 1 (
2005).
\bibitem{gas}C. Girard, R. Benoit, M. Desgagne,  Monthly Weather Rev. \bf 133\rm, 1463 (2005);
S. Kurien, V. S. L'vov, I. Procaccia and K. R. Sreenivasan, Phys.
Rev. E \bf 61\rm, 407 (2000).
%; T. S. Huang, C. W. Ho and C. J. Alexander, J. Geophys. Res.-Planets (E9) \bf 103\rm, 20267 (1998);
% A. Gluhovsky and E. Agee, J. Atmos. Sci. \bf 54\rm, 768 (1997).
\bibitem{super}F. D. M. Haldane and Y. Wu, Phys. Rev. Lett. \bf 55\rm, 2887 (1985).
%; C. Barenghi, R. Donnelly, and W. Vinen (Springer-Verlag, Berlin, 2001):
% Quantized Vortex Dynamics and superfluid Turbulence.
\bibitem{astro}
%A. Vilenkin and E. P. S. Shellard, Cosmic strings and
%other Topological Defects (Cambridge University Press, Cambridge, England, 2000);
S. Bonazzola, E. Gourgoulhon and J. A. Marck,
Phys. Rev. D \bf 56\rm, 7740 (1997); C. M. Xu, X. J. Wu and M.
Soffel, Phys. Rev. D \bf 71\rm (2005) 024030.
\bibitem{sta}A. J. Niemi, Phys. Rew. Lett. \bf 94,\ \rm  124502
(2005).
\bibitem{particle}L.Faddeev, A.J.Niemi, and U.Wiedner, hep-ph/0308240.
\bibitem{bose}A. J. Leggett, Rev. Mod. Phys. \bf 73\rm, 307 (2001).
\bibitem{crystal}I. Chuang, R. Durrer, N. Turok, and B. Yurke, Science \bf 251\rm, 1336 (1991);
M. J. Bowick, L. Chandler, E.A. Schiff, and A. M. Srivastava,
Science \bf 263\rm, 943 (1994).
\bibitem{H}E. Babaev, A. Sudb\o, and N.W.
Ashcroft, Nature (London) 431, 666 (2004).
\bibitem{HF}F. Huang and S. Y. Lou, Phys. Lett. A \bf 320\rm,  428
(2004); V. L. Saveliev, M. A. Gorokhovski, Phys. Rev. E \bf 72\rm,
016302 (2005).
%; V. Andereev, Nonl. Math. Phys. \bf 3\rm, 196 (1996).
\bibitem{group_JPA}S. Y. Lou and H. C. Ma, J. Phys. A: Math. Gen. \bf 38\rm,
L129 (2005).
\bibitem{CPL}S. Y. Lou, Chin. Phys. Lett., \bf 21\rm, 1020
(2004).
%; S. Y. Lou, C. Rogers and W. K. Schief, Stud. Appl. Math. \bf 113\rm, 353 (2004).
\bibitem{ring}S. Y. Lou, J. Math. Phys. \bf 41\rm, 6509 (2000).
%; J. Phys. A: Math. Gen., \bf 36\rm, 3877 (2003).
\bibitem{ring1}X. Y. Tang, S. Y. Lou and Y. Zhang, Phys. Rev. E. \bf 66\rm, 046601
(2002).
%; X. Y. Tang and S. Y. Lou, J. Math. Phys. \bf 44\rm, 4000 (2003).
\bibitem{typhoon}K. Emanuel, Nature, \bf 436\rm, 686 (2005).
\bibitem{eye}Q. H. Zhang, S. J. Chen, Y. H. Kuo and R. A. Anthes,
Monthly Weather Rev. \bf 133\rm, 725 (2005);
\end{thebibliography}
\end{document}